\documentclass[aps,prl,preprint,showpacs,superscriptaddress]{revtex4} 

\usepackage{bm}
\usepackage{amsmath}
\usepackage{graphicx}

\begin{document}

\title{Nonlinear propagation of broadband intense electromagnetic waves 
  in an electron--positron plasma}  
  
\author{M. Marklund}
\affiliation{Centre for Nonlinear Physics, Department of Physics,
  Ume{\aa} University, SE--90187 Ume{\aa}, Sweden}
\affiliation{Centre for Fundamental Physics, Rutherford Appleton Laboratory,
  Chilton, Didcot, Oxon OX11 OQX, U.K.}
  
\author{B. Eliasson}
\affiliation{Institut f\"ur Theoretische Physik IV and Centre for Plasma Science and
Astrophysics, Fakult\"at f\"ur Physik und Astronomie, Ruhr-Universit\"at Bochum, 
D--44780 Bochum, Germany}

\author{P. K. Shukla}
\affiliation{Institut f\"ur Theoretische Physik IV and Centre for Plasma Science and
Astrophysics, Fakult\"at f\"ur Physik und Astronomie, Ruhr-Universit\"at Bochum, D--44780 Bochum, Germany}
\affiliation{Centre for Nonlinear Physics, Department of Physics, Ume{\aa} University, 
SE--90187 Ume{\aa}, Sweden}
\affiliation{CCLRC Centre for Fundamental Physics, Rutherford Appleton Laboratory,
Chilton, Didcot, Oxon OX11 OQX, U.K.}
\affiliation{SUPA, Department of Physics, University of Strathclyde, Glasgow G4 ONG, U. K.}
\affiliation{GoLP/Centro de F\'isica de Plasmas, Instituto Superior T\'ecnico, 1096 Lisboa Codex,
Portugal}
\begin{abstract}
  A kinetic equation describing the nonlinear evolution of intense
  electromagnetic pulses in electron--positron (e-p) plasmas is presented.
  The modulational instability is analyzed for a relativistically intense 
  partially coherent pulse, and it is found that the modulational instability 
  is inhibited by the spectral pulse broadening. A numerical study for the 
  one-dimensional kinetic photon equation is presented. Computer simulations 
  reveal a Fermi-Pasta-Ulam-like recurrence phenomena for localized broadband 
  pulses. The results should be of importance in understanding the nonlinear 
  propagation of broadband intense electromagnetic pulses in e-p plasmas 
  in laser-plasma systems as well as in astrophysical plasma settings.
\end{abstract}

\maketitle

\section{I. Introduction}
Electron-positron plasmas are believed to be an important ingredient of the
early universe and in astrophysical objects such as pulsars, supernova remnants and
active galactic nuclei, and in gamma-ray bursts \cite{Piran}. In such extreme
environments, the electron-positron pairs may be created by collisions between 
particles that are accelerated by electromagnetic and electrostatic waves and/or
by gravitational forces. In pulsar environments, there is also a possibility of 
pair creation via high-energy curvature radiation photons that are triggered by
charged particles streaming along the curved magnetic field \cite{Sturrock}, with 
a resulting collection of positrons at the polar caps of the pulsar \cite{Arons,Michel}.
High-energy laser-plasma interactions and fusion devices on Earth also constitute a 
source of electron-positron plasmas. Experiments with petawatt lasers 
(with intensities exceeding $10^{20}\,\mathrm{W/cm}^2$) have demonstrated the production of MeV electrons and evidence 
of positron production via electron collisions \cite{Campbell,Cowan}.
Positrons are also believed to be created in post-disruption plasmas in large tokamaks through
collisions between MeV electrons and thermal particles \cite{Helander}.

Various collective phenomena in electron-positron plasmas have been demonstrated in 
the laboratory, such as the creation of wake fields by ultra-relativistic positron beams \cite{Blue}
and two-stream instabilities in streaming electron-positron plasmas \cite{Greaves}.
The natural saturation mechanism for a two-stream instability is particle trapping in
which electrons and positrons are trapped in the electrostatic potential of the large-amplitude
wave. Theoretical and numerical studies have shown that solitary electron/positron holes 
can exist both in a non-relativistic pair plasma \cite{Popel95,Eliasson05a,Eliasson06a} and in a 
relativistically hot electron-positron-ion plasma \cite{Eliasson05b}. Intense electromagnetic
radiation introduces a new nonlinearity in the electron-positron plasma via the 
relativistic mass increase of the plasma particles in the quivering electric field of the 
electromagnetic waves \cite{Shukla86,Bingham04}. It has been demonstrated theoretically 
that relativistic electromagnetic solitary waves can exist in a cold electron-positron plasma, 
if there is an inclusion of massive ions \cite{Berezhiani}. Several authors \cite{Stenflo,Yu}
have also considered the nonlinear propagation of small amplitude electromagnetic waves
and Alfv\'en vortices in a strongly magnetized electron-positron plasma.

Over recent years, the laser pulse intensity has been strongly increasing, and
over the next five to ten years it is expected to reach focal intensities of $\sim 10^{26}\,\mathrm{W/cm^2}$
\cite{Mourou-etal}. Thus, the interaction between plasmas and intense laser pulses is currently 
of great interest \cite{r1,r2,bingham,r3,r4}. As laser intensities keep increasing, laboratory plasma
experiments will experience the effects of the nonlinear quantum vacuum \cite{marklund-shukla},
such as pair creation and photon splitting. Theoretical studies of nonlinear laser-plasma interactions
mostly make the simplified assumption of a coherent, monochromatic wave propagation in plasmas. 
It is then relatively easy to derive analytic expressions for various instabilities (e.g. 
relativistic Raman forward and backward scattering instabilities and the modulational instability 
in an electron-ion plasma \cite{Mckinstrie92}) and the formation of envelope electromagnetic 
solitons \cite{Shukla}.  

However, the assumption about a monochromatic wave is not always true, e.g. if the electromagnetic 
waves are excited via turbulent processes. Then, we would have a distribution of waves with different 
wavelengths and frequencies that are only partially coherent.  Theories for photon acceleration of 
partially coherent waves have been developed in the framework of nonlinear optics \cite{tito} and 
are based on the Wigner's kinetic description \cite{wigner}. The latter has been used to 
investigate the nonlinear instability of random phased small amplitude electromagnetic
waves in a nonlinear dispersive medium \cite{fedele}.

In this paper, we consider the nonlinear propagation of relativistically intense broadband laser 
pulses in an electron--positron plasma. In order to analyze the properties of such partially 
coherent laser pulses, we perform a Wigner analysis of the modified nonlinear Schr\"odinger
equation, and obtain a wave kinetic equation for the photon quasi-particles. The resulting kinetic equation
is investigated both analytically and numerically. Specifically, we present the modulational instability 
and the dynamics of partially coherent intense laser pulses. It is found that the partial coherence 
of the laser pulse yields a reduced growth rate for the modulational instability.
The fully nonlinear evolution of the wave kinetic equation is investigated by means 
of numerical simulations. The latter show a complex dynamics of broadband laser pulses
in that  the laser envelope contracts and disperses in a quasi-periodic manner, somewhat 
similar to the Fermi-Pasta-Ulam recurrence phenomenon encountered for the nonlinear Schr\"odinger 
equation. The present results should be useful for understanding the nonlinear propagation of
broadband intense electromagnetic pulses through pair plasmas such as those in laser-plasma systems 
and in astrophysical settings.    

\section{II. Governing equations and the modulational instability}

Let us consider the nonlinear propagation of intense circularly polarized electromagnetic 
waves in pair plasmas without ions. Accounting for the relativistic mass increase of the
pairs and quasi-stationary density fluctuations driven by the relativistic ponderomotive
force, Shukla {\it et al.} \cite{shukla-marklund-eliasson} derived the fully nonlinear 
equation
\begin{equation}\label{eq:nlse}
  i\frac{\partial a}{\partial t} + \frac{1}{2}\nabla^2a 
    + \left[ 1 - \frac{\exp[\beta(1 - \sqrt{1 + |a|^2}\,)]}{\sqrt{1 + |a|^2}} \right]a = 0 ,
\end{equation}
which shows the evolution of the normalized vector potential in pair plasmas. Here $a = eA/mc^2$ is the normalized vector potential, $A$ is the amplitude of the circularly polarized vector potential, $m$ is the electron mass, $e$ is the magnitude of the electron charge, $c$ is the speed of light in vacuum, $\beta = 2\beta_e\beta_p/(\beta_e + \beta_p)$ is the dimensionless temperature parameter with $\beta_{e,p} = (c/v_{Te,p})^2$, $v_{Te,p} = (T_{e,p}/m)^{1/2}$ is the thermal speed, $T_{e}$ ($T_{p}$) is the electron (positron) temperature, we have normalized the time and co-moving spatial variables by $\omega_0/\omega_{\mathrm{p}}^2$ and $c/\omega_{\mathrm{p}}$, respectively, $\omega_{\mathrm{p}} = (4\pi n_0e^2/m)^{1/2}$ is the electron plasma frequency, $n_0$ is the unperturbed electron density, and $\omega_0$ is the central wave frequency. 
Numerical analysis of Eq.\ (1) revealed that weakly modulated electromagnetic (em) pulses 
would undergo collapse, leading to strong intensification of the localized em pulses.

Even though Eq.\ (\ref{eq:nlse}) contains a saturation nonlinearity, 
halting of collapse could be obtained by spectral broadening techniques,
well-known in inertial confinement fusion. Moreover, in many applications
of such an equation, for example astrophysical systems, the pulses can
be partially coherent. Thus, understanding the dynamics and stability
of electromagnetic pulses taking incoherence effects into account
may be important for practical purposes. 

A canonical way for analyzing effects of partial coherence is to use
the so called Wigner function, defined as the Fourier transform of the 
two-point correlation function of the vector
potential $a$ according to \cite{wigner,tito}
\begin{equation}\label{eq:wigner}
  \rho(t,\mathbf{r},\mathbf{k}) = \frac{1}{(2\pi)^3}\int\,d\bm{\xi}\,
    e^{i\mathbf{k}\cdot\bm{\xi}}\langle
      a^*(t,\mathbf{r} + \bm{\xi}/2)a(t,\mathbf{r} - \bm{\xi}/2)
    \rangle ,
\end{equation}
where the angular bracket denotes the ensemble average. 
The Wigner function represents a generalized distribution function
for quasi-particles, in this case photons. From the definition (\ref{eq:wigner}),
one finds the relation
\begin{equation}\label{eq:intensity}
  I(t,\mathbf{r}) = \int d\mathbf{k}\,\rho(t,\mathbf{r},\mathbf{k}) ,
\end{equation}
where $I = \langle|a|^2\rangle$. Applying the time derivative to the
definition (\ref{eq:wigner}) and using Eq.\ (\ref{eq:nlse}), we obtain
the kinetic equation
\begin{equation}\label{eq:kinetic}
  \frac{\partial\rho}{\partial t} + \mathbf{k}\cdot\bm{\nabla}\rho 
    - \frac{2\exp[\beta(1 - \sqrt{1 + I}\,)]}{\sqrt{1 + I}}
    \sin\left( \frac{1}{2}\stackrel{\leftarrow}{\bm{\nabla}}
    \cdot\stackrel{\rightarrow}{\bm{\nabla}}_k 
    \right)\rho = 0 ,
\end{equation}
for the quasi-particles. Here the $\sin$-operator is defined in terms of its Taylor
expansion. Keeping only the first term in the latter, which corresponds to the
long wavelength limit, we obtain a photon kinetic (or the Liouville) equation
\begin{equation}
  \frac{\partial\rho}{\partial t} + \mathbf{k}\cdot\bm{\nabla}\rho 
    - \bm{\nabla}\left[\frac{\exp[\beta(1 - \sqrt{1 + I}\,)]}{\sqrt{1 + I}}\right]\cdot
    \bm{\nabla}_k\rho = 0 .
\end{equation}
However, the photon kinetic limit is only valid for weak spatial variations in $\rho$, 
and produces a modulational instability growth rate which is unbounded as the wavenumber 
of the photons increases. Thus, for a broad spatial spectral distribution of photons 
the dynamics of the Liouville equation is highly unstable and strongly deviates
from the full Wigner dynamics, and will therefore not be analyzed further here.    

Equations (\ref{eq:intensity}) and (\ref{eq:kinetic}) constitute a complete description
of partially coherent nonlinear photons in an electron--positron plasma.

Next, we perform a perturbation analysis of Eq.\ (\ref{eq:kinetic}). Letting
$\rho = \rho_0(\mathbf{k}) + \rho_1\exp(i\mathbf{K}\cdot\mathbf{r} - i\Omega t)$,
where $|\rho_i| \ll \rho_0$, we linearize Eqs.\ (\ref{eq:intensity}) and
(\ref{eq:kinetic}) with respect to the perturbation variables. From Eq.\ (\ref{eq:kinetic})
we then obtain
\begin{equation}
    -i\Omega\rho_1 + i\mathbf{k}\cdot\mathbf{K}\rho_1 
    -2I_1\frac{dU(I_0)}{dI_0}
    \sin\left( \frac{i}{2}\mathbf{K}\cdot{\bm{\nabla}}_k 
    \right)\rho_0 = 0 ,
\end{equation}
where 
\begin{equation}
  U(I_0) = \frac{\exp[\beta(1 - \sqrt{1 + I_0}\,)]}{\sqrt{1 + I_0}}.
\end{equation}
Combining (5) with Eq.\ (\ref{eq:intensity}) we obtain the nonlinear dispersion relation
\begin{equation}\label{eq:disprel}
  1 = \frac{dU(I_0)}{dI_0}\int d\mathbf{k}\frac{\rho_0(\mathbf{k} - \mathbf{K}/2) 
    - \rho_0(\mathbf{k} + \mathbf{K}/2)}{ \Omega - \mathbf{k}\cdot\mathbf{K}} ,
\end{equation}
which is valid for partially coherent intense laser pulses in an electron--positron plasma.

In the case of a monochromatic pulse, the background distribution function
satisfies $\rho_0(\mathbf{k}) = I_0\delta(\mathbf{k} - \mathbf{k}_0)$ for some 
wavevector $\mathbf{k}_0$. The dispersion relation (\ref{eq:disprel}) then reads
\begin{equation}\label{eq:coherent}
  \Omega = \mathbf{K}\cdot\mathbf{k}_0 \pm 
    \left[  \frac{1}{4}K^4 + I_0\frac{dU(I_0)}{dI_0}K^2 \right]^{1/2} ,
\end{equation}
which is agreement with the results found in Ref.\ \cite{shukla-marklund-eliasson}. 
The growth rate $\Gamma = -i\Omega$ is given by 
\begin{equation}\label{eq:coherent2}
  \Gamma = \left[ \frac{I_0}{2}\left( \beta + \frac{U(I_0)}{1 + I_0} \right)K^2 
  - \frac{1}{4}K^4  \right]^{1/2}.
\end{equation}

In order to simplify the perturbation analysis, we introduce the assumption of  
one-dimensional partially coherent photon propagation along the $z$-axis. 
We investigate the case of partial coherence using the background Lorentz
distribution
\begin{equation}\label{eq:lorentz}
  \rho_0(k) = \frac{I_0}{\pi}\frac{\Delta}{(k - k_0)^2 + \Delta^2},
\end{equation}
where $k = k_z$. The distribution (\ref{eq:lorentz}) corresponds to a
partially coherent phase of the vector potential $a_0$, giving
rise to a spectral broadening of $\rho_0$ with a width $\Delta$. From
the dispersion relation (\ref{eq:disprel}) we then obtain
\begin{equation}
  \Omega = -iK\Delta  + K k_0 \pm 
    \left[  \frac{1}{4}K^4 + I_0\frac{dU(I_0)}{dI_0}K^2 \right]^{1/2} ,
\end{equation}
thus giving (\ref{eq:coherent}) in the limit $\Delta \rightarrow 0$. We see that 
the effect of the spectral broadening is to reduce the growth rate 
according to
\begin{equation}\label{eq:incoherent}
  \Gamma = -K\Delta +  
  \left[ \frac{I_0}{2}\left( \beta + \frac{U(I_0)}{1 + I_0} \right)K^2 
  - \frac{1}{4}K^4  \right]^{1/2} .
\end{equation}
We note that as $\Delta \rightarrow 0$, we retrieve the expression (\ref{eq:coherent2}).

A comparison between the coherent and incoherent modulational instability growth rates 
is presented in Fig.\ 1. We note from Eq. (\ref{eq:incoherent}) that the growth rate is 
larger than zero only if $\Delta < (I_0/2)^{1/2}[\beta+U(I_0)/(1+I_0)]^{1/2}$, and  in the
small amplitude limit $I_0\ll 1$ we have the condition 
$\Delta < (I_0\beta/2)^{1/2}$ for the modulational instability. The Liouville 
equation is obtained from the Wigner's kinetic equation when the product
of the spatial length scale $L$ and the spectral width $\Delta$ is large enough. For ultra-short laser pulses with a focal intensities of $\mathcal{I} \approx \omega_0^2|A|^2/c \sim 10^{21}\,\mathrm{W/cm^2}$, wavelength $\lambda_0 \sim 800\,\mathrm{nm}$, and spectral width $\delta\lambda\sim 50\,\mathrm{nm}$ (relevant for e.g.\ the Astra laser and its upgrade Astra Gemini at the Rutherford Appleton Laboratory, Oxfordshire, U.K.), we have the coherence length $\ell_c = \lambda_0^2/\delta\lambda \approx 13\,\mathrm{\mu m}$. Since $\Delta \sim c/\ell_c\omega_p$ we have the modulational instability criteria $(n_0/n_\mathrm{crit})(c/v_{Te})^2 > m^2c^5/4\pi e^2\mathcal{I}\ell_c^2$, where $n_{\mathrm{crit}} = m\omega_0^2/4\pi e^2$. For the above intensity the criteria for the modulational instability to take place then becomes $2.1\times 10^{-6} < (n_0/n_\mathrm{crit})(c/v_{Te})^2$, and if the electron temperature is $T_e \sim 5\,\mathrm{keV}$, we find that $n_0 > 4\times 10^{-22}n_{\mathrm{crit}} \sim 10^{-2}\,\mathrm{cm}^{-2}$.

\section{III. Nonlinear dynamics of broadband intense electromagnetic pulses}

In order to understand the long term behavior of modulationally unstable 
broadband intense electromagnetic pulses in pair plasmas, we carry out 
the numerical analysis of the kinetic photon equation (\ref{eq:kinetic}) 
in the one-dimensional case. Accordingly, we numerically solve the set of equations
\begin{equation}\label{eq:1Dkinetic}
  \frac{\partial\rho}{\partial t} + k_x\frac{\partial \rho}{\partial x} 
    - \frac{2\exp[\beta(1 - \sqrt{1 + I}\,)]}{\sqrt{1 + I}}
    \sin\left( \frac{1}{2}\stackrel{\leftarrow}{\frac{\partial}{\partial x}}\cdot
    \stackrel{\rightarrow}{\frac{\partial}{\partial k_x}} 
    \right)\rho = 0 ,
\end{equation}
and 
\begin{equation}\label{eq:1Dintensity}
  I(t,x) = \int_{-\infty}^{\infty} dk_x\,\rho(t,x,k_x).
\end{equation}

For the numerical solutions of Eqs. (\ref{eq:1Dkinetic}) and (\ref{eq:1Dintensity}), 
we use a Fourier method for the Vlasov-Poisson system of equations \cite{Eliasson01,Eliasson06b}, 
which is slightly modified to solve the Wigner equation. For this purpose, we use
the Fourier transform pair
\begin{eqnarray}
  &&\rho(t,x,k_x)=\int_{-\infty}^{\infty}d\eta \widehat{\rho}(t,x,\eta)e^{-i\eta k_x},
  \\
  &&\widehat{\rho}(t,x,\eta)=\frac{1}{2\pi}\int_{-\infty}^{\infty}dk_x \rho(t,x,k_x)e^{i\eta k_x},
\end{eqnarray}
to obtain the Fourier transformed version of Eqs. (\ref{eq:1Dkinetic}) and (\ref{eq:1Dintensity}),
\begin{equation}\label{eq:1DFkinetic}
  \frac{\partial\widehat{\rho}}{\partial t} - i \frac{\partial^2 \widehat{\rho}}{\partial \eta \partial x} 
    + \widehat{\rho} \sin\left( \frac{i\eta }{2}\frac{\partial}{\partial x} 
    \right)\left[\frac{2\exp[\beta(1 - \sqrt{1 + I}\,)]}{\sqrt{1 + I}}\right]
    = 0 ,
\end{equation}
and 
\begin{equation}\label{eq:1DFintensity}
  I(t,x) = 2\pi\widehat{\rho}(t,x,\eta)_{\eta=0},
\end{equation}
respectively. The Fourier transformed system of Eqs. (\ref{eq:1DFkinetic})
and (\ref{eq:1DFintensity}), which is equivalent to 
Eqs. (\ref{eq:1Dkinetic}) and (\ref{eq:1Dintensity}), is solved numerically
in a periodic box in $x$ space, where the $x$ derivatives are approximated
with a pseudo-spectral method. In this method, the differential operator
$\partial/\partial x$ turns into a multiplication by $iK_j$ and the operator 
$\sin[({i\eta }/{2}){\partial}/{\partial x}]$ turns into a multiplication
by $\sin(-{\eta K_j}/{2})$, where $K_j=2\pi j/L$ is the
spatial wavenumber, $j=0,\,\pm 1,\,\pm2\ldots$ and $L$ is the length of 
the spatial domain. We note that the integral over $k_x$ in Eq. (\ref{eq:1Dintensity})
is transformed into a simple evaluation of $\widehat{\rho}$ at $\eta=0$ in Eq. (\ref{eq:1DFintensity}).
As an initial condition for the simulation, we take the Fourier transformed 
Lorentz distribution (\ref{eq:lorentz})
\begin{equation}
  \widehat{\rho}_0(\eta)=\frac{I_0}{2\pi}\exp(-\Delta|\eta|+ik_0\eta),
\end{equation}
and the same parameters as in Fig.\ 1. In Figs.\ 2 and 3, we use $I_0=1/4$, 
$\Delta=0.1$, and $k_0=0.1$ as initial conditions. 
A small amplitude noise (random numbers of order $10^{-3}$) is added to the
initial condition to give a seed for the modulational instability.

We display the temporal evolution of the em wave intensity $I$ in Fig.\ 2. Here we see
an initially linear growth phase and a wave collapse that takes place around $t=90$. The
wavelength of the collapsing pulse is approximately $\lambda=10$, and thus the most unstable
mode is $k^*=2\pi/10\approx 0.6$, in good agreement with the theoretical prediction in Fig.\ 1.
After the initial pulse collapse, the solution shows a chaotic behavior, where some localized
envelopes show a semi-periodic behavior, and we also see a sequence of merging and splitting
of wave groups. This behavior is somewhat similar to the Fermi-Pasta-Ulam recurrence phenomenon
\cite{campbell-etal}, which has been observed in nonlinear optics experiments \cite{simaeys} 
and has been explained theoretically in the framework of the nonlinear Schr\"odinger 
equation \cite{infeld,akhmediev}.
In Fig.\ 3, we show Wigner's distribution function $\rho$ at different times. (It is obtained by 
numerically inverse Fourier transforming $\widehat{\rho}$ to obtain the real-valued $\rho$.) 
We note that the distribution function shows a complex dynamics and takes both positive and
negative values, in contrast to solutions of the Vlasov equation for particles that takes only 
positive values.  In Figs.\ 4 and 5, we repeated the simulation with the larger spectral 
width $\Delta=0.3$ in the initial conditions. We used the same intensity $I_0=1/4$, as in the first 
simulation, and we used $k_0=0.05$. For this case, the temporal evolution of the pulse intensity,
as shown in Fig.\ 4, exhibits a slower growth rate (compared to Fig.\ 2 with the smaller $\Delta=0.1$), 
and nonlinearly collapsing wave envelopes have smaller intensity maxima. The snap shots of the Wigner
distribution function, depicted in Fig.\ 5, shows the initial instability and merging of nonlinear structures
corresponding to merging wave envelopes in Fig.\ 4. Thus, a larger spectral width $\Delta$ leads to
a slower and less violent dynamics of the wave envelopes that also have larger scale sizes than for
the smaller values of $\Delta$. We recall that the theoretical treatment predicts that there
is a largest $\Delta$ above which the modulational instability vanishes, and we would instead have 
damping of waves similar as Landau damping for the Vlasov equation.

\section{IV. Summary}

In summary, we have investigated the modulational instability and the nonlinear dynamics 
of partially coherent intense electromagnetic (em) waves in an electron-positron plasma,
taking into account the relativistic mass increase of the pairs as well as large 
scale density fluctuations that are created by the relativistic ponderomotive force
of em waves. The dynamics of broadband intense em waves is governed by a photon kinetic 
(or the Liouville) equation. The latter is analyzed to obtain a nonlinear dispersion
relation, which admits the modulational instability growth rate for a Lorentzian 
distribution of partially coherent intense em waves. It is found that a broadband 
of em waves leads to a reduction of the growth rate. Furthermore, the photon kinetic equation
has been  numerically solved to understand the nonlinear dynamics of modulationally 
unstable broadband intense em pulses. Our simulation results reveal the formation of 
localized em wave packets that show some similarity to the Fermi-Pasta-Ulam recurrence 
phenomenon. For larger values of the spectral width, the system shows a dynamics on a 
slower timescale and on larger length scales, in agreement with the linear theory.
The present results should help to understand the nonlinear propagation of broadband
intense electromagnetic waves in pair plasmas that appear in inertial confinement 
fusion schemes as well as in pulsar magnetosphere and supernovae remnants. 

\acknowledgments

This work was partially supported by the Swedish Research Council and the 
Deutsche Forschungsgemeinschaft.

\newpage

\noindent Figure 1: The modulational instability growth rates plotted
  as a function of $K$, as given by the expression (\ref{eq:incoherent}).
  We have used $I_0=1/4$ and $\beta = 1$. The thick curve
  represents the coherent case with $\Delta =0$, the thin curve
  has a finite spectral width $\Delta$ of $0.1$. The decreased growth rate
  due to spectral broadening can clearly be seen. \\ 
  
\noindent Figure 2 (Color online): The intensity $I$ as a function of $x$ and $t$. The parameters used in
   initial are condition $I=I_0=1/4$, $\Delta=0.1$ and $k_0=0.1$. \\
   
\noindent Figure 3 (Color inline): The Wigner distribution function $\rho$ as a function
  of $x$ and $k_x$, at times $t=0$, $t=85$, $t=100$ and $t=150$
  (upper to lower panels), with the corresponding intensity in Fig. 2. 
  For the initial conditions we used $I=I_0=1/4$, $\Delta=0.1$ and $k_0=0.1$. \\
  
\noindent Figure 4 (Color online): The intensity $I$ as a function of $x$ and $t$. The parameters
  used in the initial
  condition are $I=I_0=1/4$, $\Delta=0.3$ and $k_0=0.05$. \\
  
\noindent Figure 5 (Color online): The Wigner distribution function $\rho$ as a function
  of $x$ and $k_x$, at times $t=0$, $t=300$, $t=400$ and $t=500$
  (upper to lower panels), with the corresponding intensity in Fig. 4. 
  For the initial conditions we used 
  $I=I_0=1/4$, $\Delta=0.3$ and $k_0=0.05$. 
  
\newpage 

\begin{figure}
  \includegraphics[width=0.9\columnwidth]{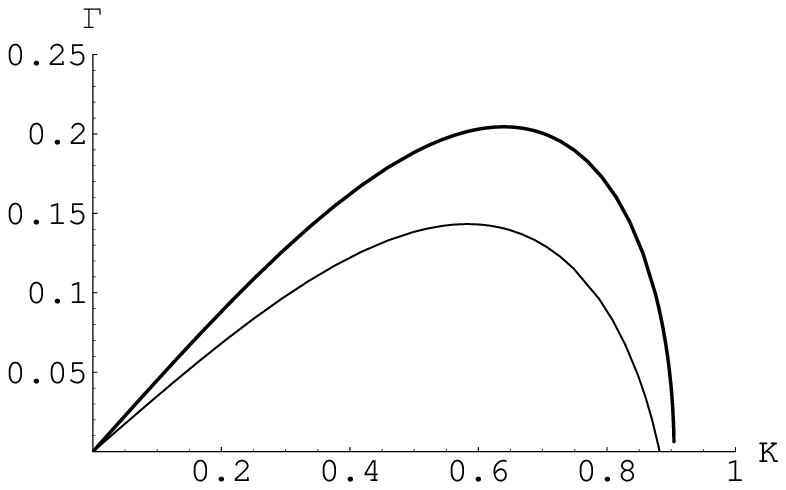}
  \caption{}
\end{figure}

\begin{figure}
  \includegraphics[width=0.8\columnwidth]{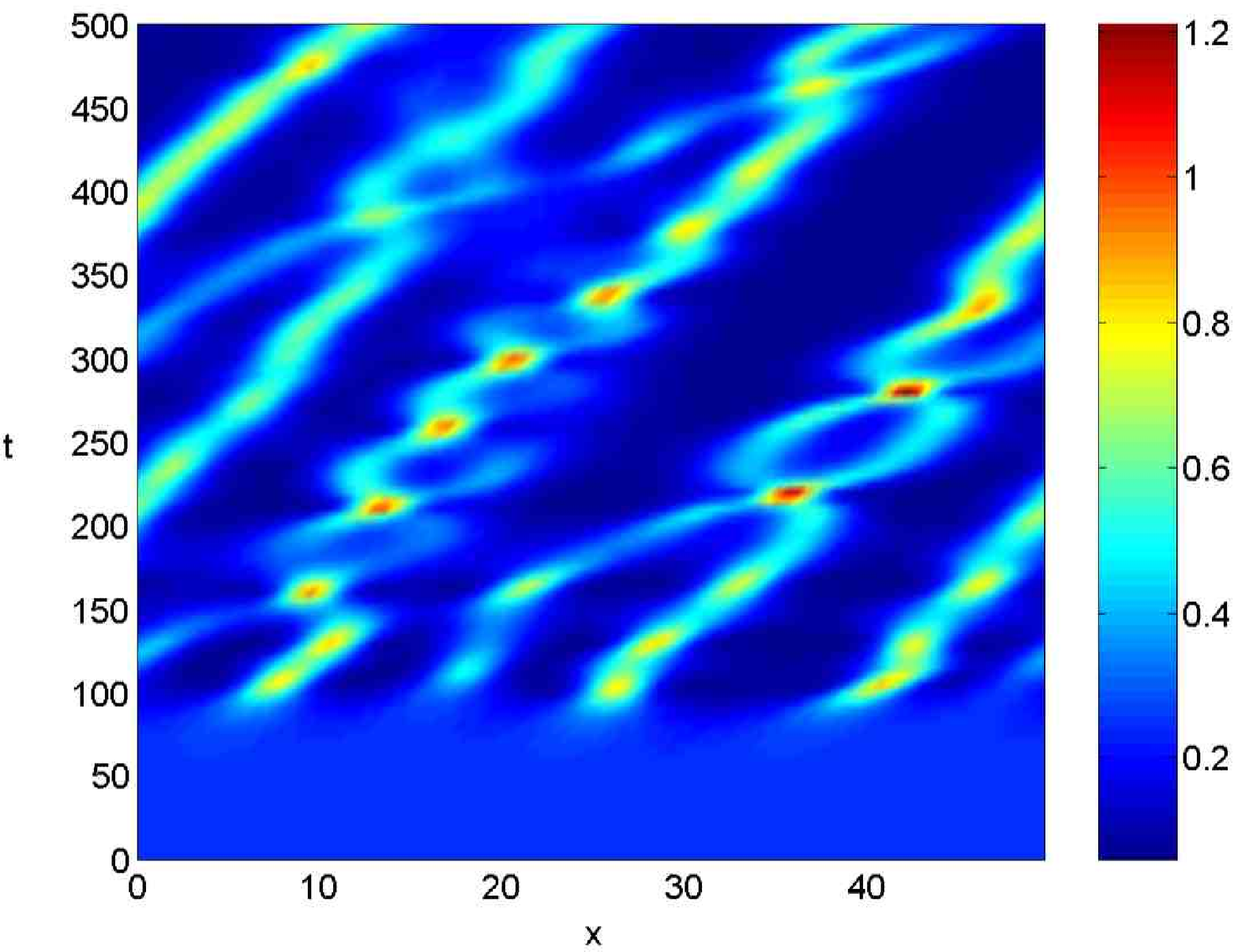}
  \caption{
  }
\end{figure}

\begin{figure}
  \includegraphics[width=0.8\columnwidth]{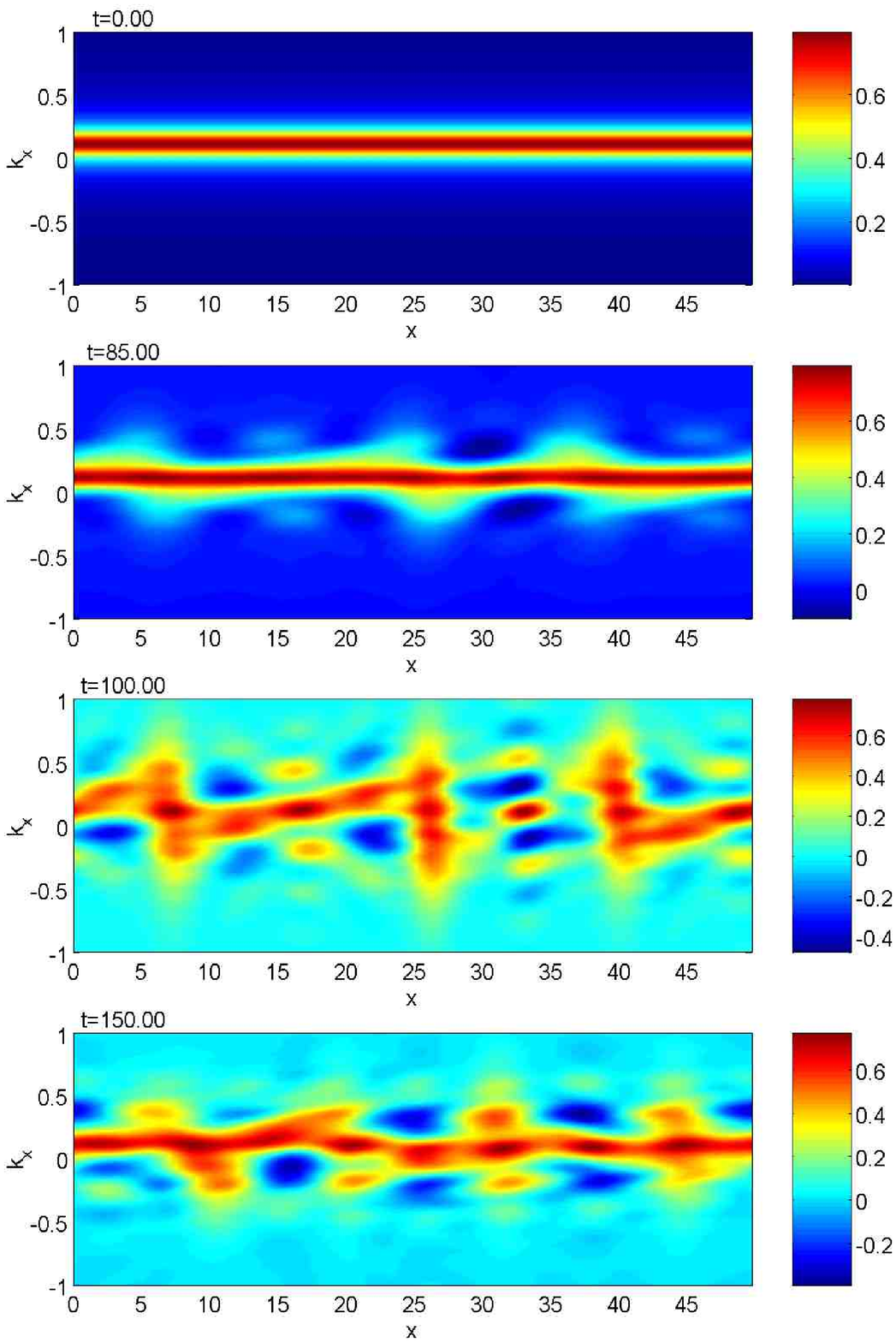}
  \caption{
  }
\end{figure}

\begin{figure}
  \includegraphics[width=0.8\columnwidth]{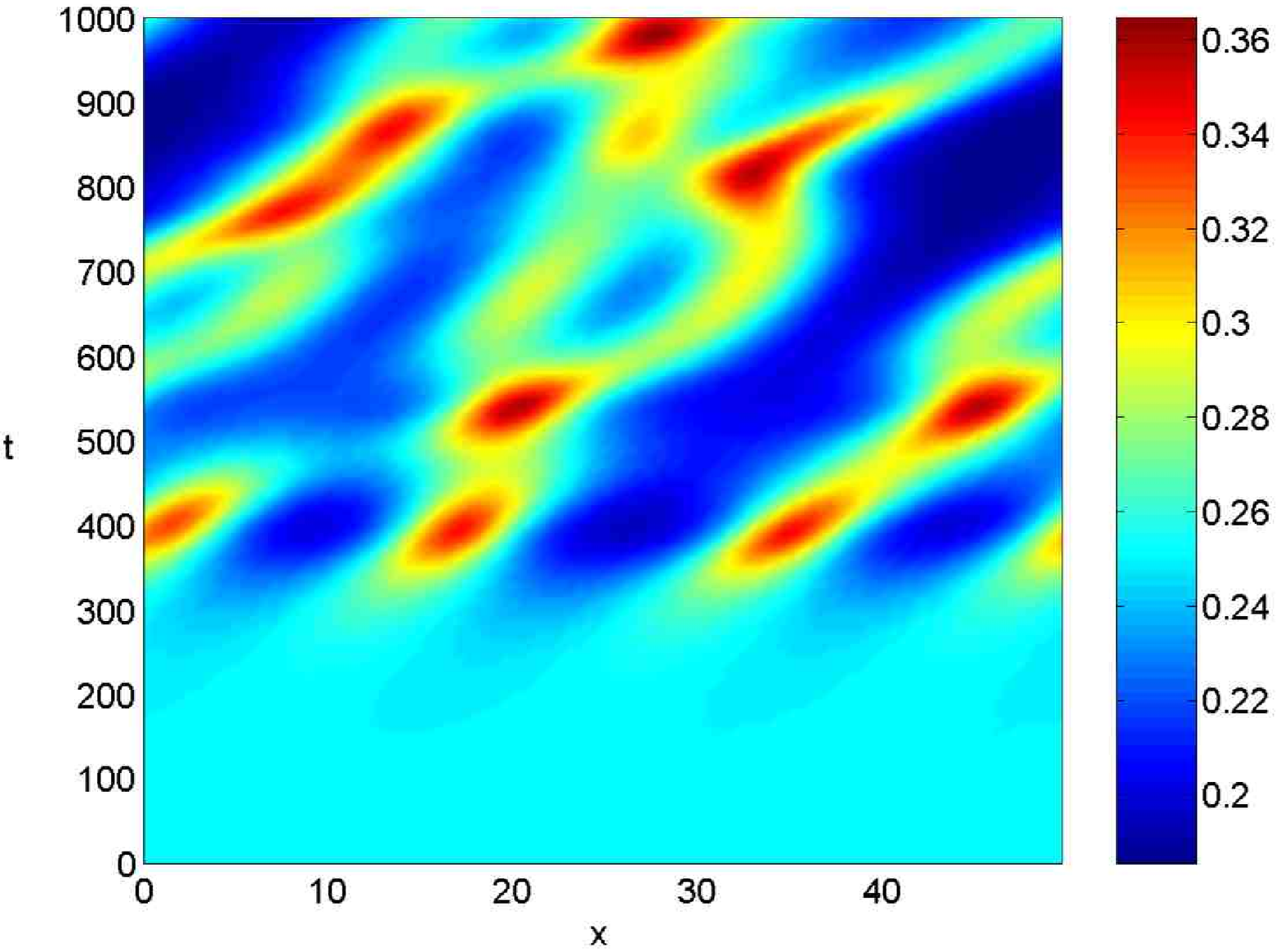}
  \caption{
  }
\end{figure}

\begin{figure}
  \includegraphics[width=0.8\columnwidth]{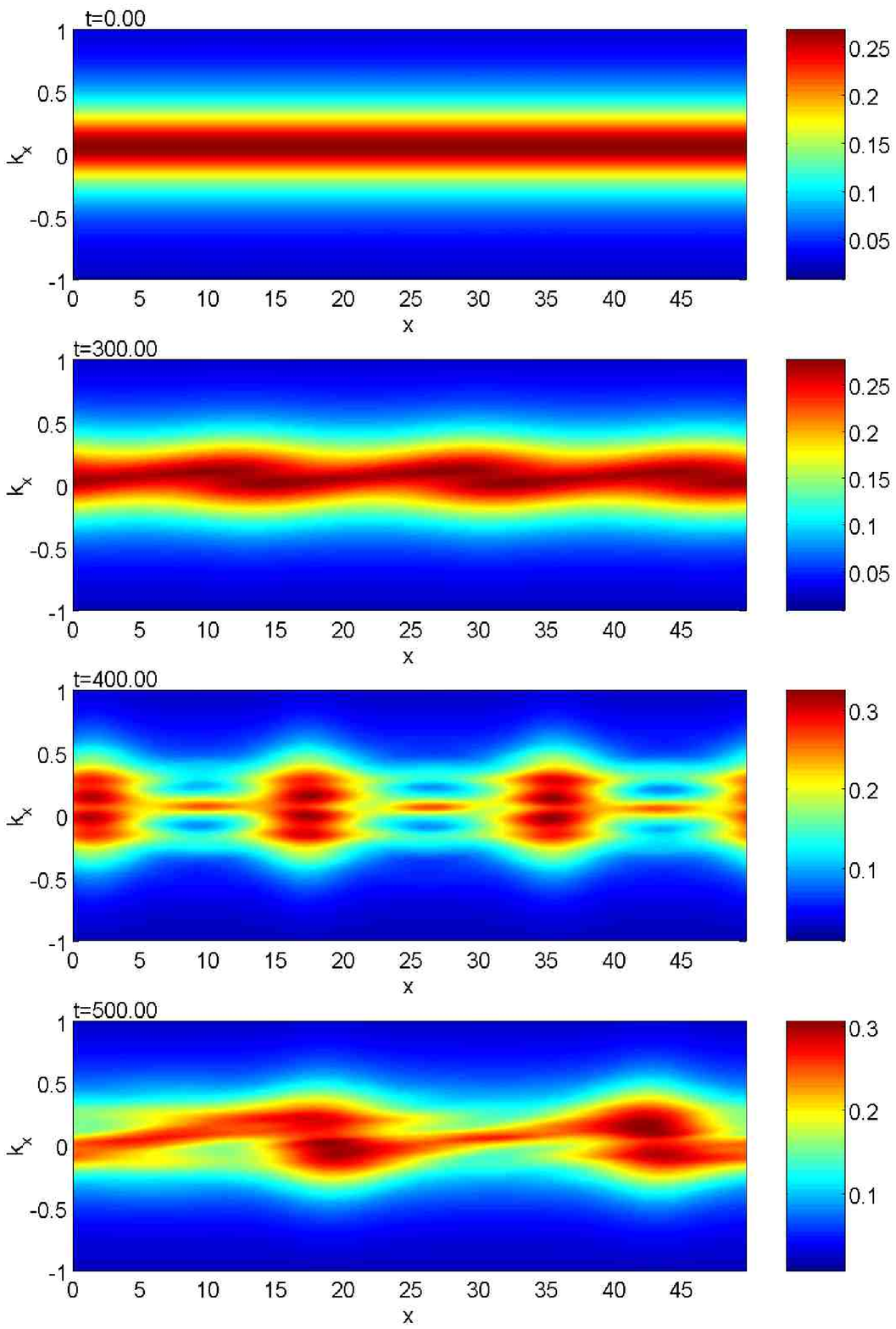}
  \caption{
  }
\end{figure}

\end{document}